\documentclass[prb,aps,twocolumn,superscriptaddress,10pt,amsmath,amssymb,nofootinbib,floatfix]{revtex4-1}
\pdfoutput=1
\usepackage{epsfig}
\usepackage{amsmath}
\usepackage{amssymb}
\usepackage{graphicx}
\usepackage{enumitem}
\usepackage{longtable}
\usepackage{dcolumn}
\makeatletter 

\begin{document}

\title{Ferroelectric/paraelectric superlattices for energy storage}

\author{Hugo Aramberri}
\affiliation{Materials Research and Technology
  Department, Luxembourg Institute of Science and Technology, 5 avenue
  des Hauts-Fourneaux, L-4362 Esch/Alzette, Luxembourg}
\affiliation{Inter-institutional Research Group Uni.lu–LIST on Ferroic Materials,
41 rue du Brill, L-4422 Belvaux, Luxembourg}

\author{Natalya S. Fedorova}
\affiliation{Materials Research and Technology
  Department, Luxembourg Institute of Science and Technology, 5 avenue
  des Hauts-Fourneaux, L-4362 Esch/Alzette, Luxembourg}
\affiliation{Inter-institutional Research Group Uni.lu–LIST on Ferroic Materials,
41 rue du Brill, L-4422 Belvaux, Luxembourg}

\author{Jorge \'I\~niguez}
\affiliation{Materials Research and Technology
  Department, Luxembourg Institute of Science and Technology, 5 avenue
  des Hauts-Fourneaux, L-4362 Esch/Alzette, Luxembourg}
\affiliation{Inter-institutional Research Group Uni.lu–LIST on Ferroic Materials,
41 rue du Brill, L-4422 Belvaux, Luxembourg}
\affiliation{Department of Physics and Materials
  Science, University of Luxembourg, 41 Rue du Brill, L-4422 Belvaux,
  Luxembourg}

\date{\today}
\begin{abstract}
The polarization response of antiferroelectrics to electric fields is
such that the materials can store large energy densities, which makes
them promising candidates for energy storage applications in
pulsed-power technologies. However, relatively few materials of this
kind are known. Here we consider ferroelectric/paraelectric
superlattices as artificial electrostatically-engineered
antiferroelectrics. Specifically, using high-throughput
second-principles calculations, we engineer PbTiO$_{3}$/SrTiO$_{3}$
superlattices to optimize their energy-storage performance at room
temperature (to maximize density and release efficiency) with respect
to different design variables (layer thicknesses, epitaxial
conditions, stiffness of the dielectric layer). We obtain results
competitive with the state-of-the-art antiferroelectric capacitors and
reveal the mechanisms responsible for the optimal properties.
\end{abstract}

\maketitle

%\section*{Introduction}
\vspace{5mm}{\bf Introduction}

\begin{figure}
  \includegraphics[scale=0.20]{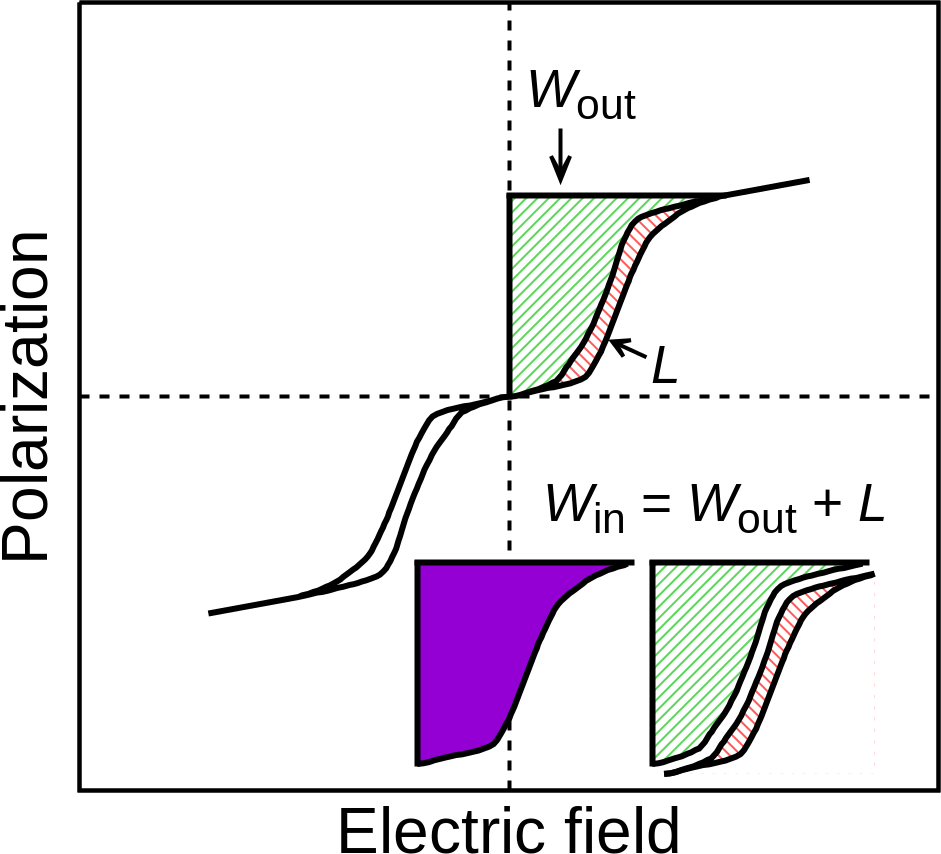}
	\caption{Energy storage in $P$-$\varepsilon$ loops. The energy density required to charge the system ($W_{\mathrm{in}}$) is equal to the recovered energy density
	upon discharge ($W_{\mathrm{out}}$) plus the loss ($L$). Energy densities are proportional to areas in $P$-$\varepsilon$ diagrams.}
	\label{fig_explan}
\end{figure}

One of the limiting factors in the miniaturization of present day
electronics is the relatively large size of their capacitors, due to
their somewhat low energy density. Antiferroelectric materials could
help solve this problem~\cite{zhu12,rabe13}.  These compounds present
an antipolar structure of electric dipoles, yielding overall no net
polarization $P$. Yet, applying a large enough electric field
$\varepsilon$ can switch the system onto a polar state of large
polarization (see Fig.~\ref{fig_explan}). The energy density required
to charge the system in this way ($W_{\rm{in}}$) is proportional to
the area to the left of the charging branch in the $P$-$\varepsilon$
loop, which is indicated in purple in Fig.~\ref{fig_explan}. Upon
removal of the electric field the released energy density is
proportional to the area to the left of the discharge branch
($W_{\rm{out}}$, in green in the figure). The energy loss ($L$) is
therefore proportional to the area enclosed by the loop in the
$P$-$\varepsilon$ diagram (red-shaded area in
Fig.~\ref{fig_explan}). This field-driven phase transition can be
utilized in a capacitor with possible applications in pulsed power
technologies~\cite{burn72}. Still, relatively few
antiferroelectrics are known~\cite{rabe13}, which hampers the
optimization of the effect. Hence, there is a pressing need to
discover new antiferroelectric materials.

In the past years, several efforts have been devoted to improving the
energy-storing performance of known antiferroelectrics. Polymers and
ceramic/polymer composites can present high breakdown fields, but
store modest energy densities and typically suffer from poor thermal
stability~\cite{gong16,li20b}. Several works have reported noticeable
energy densities in samples of hafnia- and zirconia-based
antiferroelectrics~\cite{park14,pevsic16,zhang17,kim19,yi21}, although
with modest efficiency $W_{\rm{out}}/W_{\rm{in}}$ due to a strong
hysteretic behaviour. A more promising route explored by many groups
is chemical substitution in antiferroelectric
perovskites~\cite{ma09a,ma09b,ye11,yao11,hao11,hu14,peng15,ahn15,xu17,pan19a,kim20,li20a}. To
our best knowledge, the highest energy density in an antiferroelectric
experimentally achieved to date is 154~J~cm$^{-3}$ and corresponds to
a complex perovskite solid solution~\cite{peng15}.

Several works have found or predicted antiferroelectricity in
electrostatically frustrated perovskite oxides. Antiferroelectric
phases were measured in KNbO$_{3}$/KTaO$_{3}$~\cite{sigman02} and
SrTiO$_3$/BaZrO$_3$~\cite{christen03} superlattices, though the former
present scant thermal stability, and the latter display
antiferroelectricity for very thin layers. Theoretical works have
predicted the appearance of antipolar states in BaTiO$_{3}$/BaO
superlattices~\cite{bousquet10} (though the electric field response
was not computed) or electrostatically engineered ferroelectric thin
films~\cite{glazkova14} (which display small stability
regions). Still, the measured or predicted stability windows in these
systems are rather narrow, leaving little room for optimization.

In the last years, multidomain structures have been reported in
PbTiO$_{3}$/SrTiO$_{3}$
superlattices~\cite{zubko10,yadav16,das19}. The observed dipole
structures can be deemed antipolar, and hence these systems are good
candidates to display antiferroelectric-like behaviour. These
superlattices have attracted attention lately since they have been
found to host negative capacitance~\cite{zubko16}, non-trivial dipole
topologies~\cite{gonccalves19,das19,abid21}, and subterahertz
collective dynamics~\cite{li21}, with possible applications for
voltage amplification and in electric-field-driven data processing,
among others.

In this work we test the performance of ferroelectric/paraelectric
superlattices as artificial antiferroelectrics for energy storage,
taking PbTiO$_{3}$/SrTiO$_{3}$ as a relevant model system. We show
that the antipolar multidomain state of these heterostructures can be
switched to a monodomain polar one under an electric field, yielding
response curves similar to that of Figure~\ref{fig_explan}, and thus
displaying antiferroelectric-like behaviour.  These superlattices
offer multiple design variables, including the PbTiO$_{3}$ and
SrTiO$_{3}$ layer thicknesses, the epitaxial strain imposed by a
substrate, or the stiffness of the dielectric layer (which can be
controlled through its composition), which are expected to have a
significant impact in their electrostatic response and thus in its
performance as capacitors. We explore these optimization possibilities
by using second-principles simulation methods (which have proven
successful to describe these superlattices in previous
works~\cite{zubko16,das19,yadav19}) to run a high-throughput
investigation over these design variables.  We also reveal the
underlying physics yielding the best properties.

%\section*{Results}
\vspace{5mm}{\bf Results}

\begin{figure}
  \includegraphics[scale=0.10]{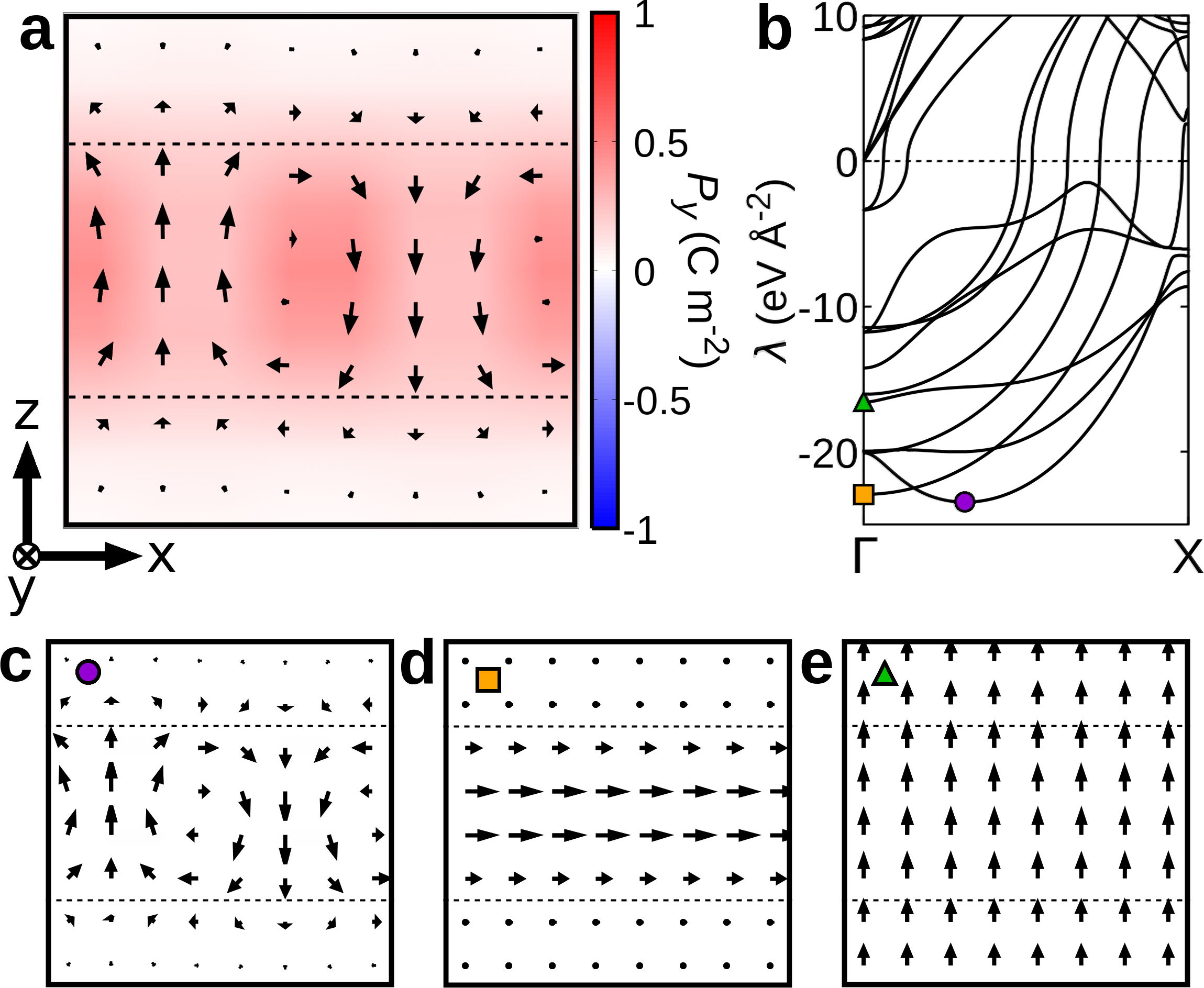}
	\caption{{\bf a} Lowest energy state of the (PbTiO$_3$)$_{4}$/(SrTiO$_{3}$)$_{4}$ superlattice. {\bf b} Phonon instabilities of the high-symmetry (PbTiO$_3$)$_{4}$/(SrTiO$_{3}$)$_{4}$ superlattice (in which the atoms in the PbTiO$_{3}$ and SrTiO$_{3}$ layers are in the cubic phase) along the $\Gamma-X$ direction. The eigenvectors of the leading instability (purple circle) and two relevant polar instabilities (orange square and green triangle) are shown in the panels below. 
		 {\bf c}, {\bf d}, and {\bf e} depict the eigenvectors for the phonons marked with the corresponding symbols in {\bf b}. In panels {\bf a}, {\bf c}, {\bf d} and {\bf e} the arrows indicate the atomic dipoles, and the out of screen component of the dipoles is colour coded according to the scale shown in {\bf a}.}
	\label{fig_gs}
\end{figure}

In ferroelectric/paraelectric superlattices with the polarization easy
axis along the growth direction, the development of a homogeneous
polar state in the ferroelectric layer is hindered by the
electrostatic penalty due to the surrounding dielectric layers. This
often results in the ferroelectric component breaking into domains
with opposite local polarizations, as shown in
Figure~\ref{fig_gs}. The resulting state can be regarded as antipolar.

Let us consider (PbTiO$_3$)$_{m}$/(SrTiO$_{3}$)$_{n}$ superlattices,
where $m$ and $n$ are the thicknesses (in perovskite unit cells) of
the PbTiO$_{3}$ and SrTiO$_{3}$ layers, respectively, and let us take
the (PbTiO$_3$)$_{4}$/(SrTiO$_{3}$)$_{4}$ system as a representative
case for the following discussion.  In Figure~\ref{fig_gs}{\bf a} we
show the lowest energy dipole configuration obtained for this material
using second principles. It presents domains with polarization along
the growth direction ($z$), yielding an overall antipolar
structure. The phonon spectrum (Figure~\ref{fig_gs}{\bf b}) of the
high-symmetry superlattice state (in which all the atoms are fixed to
the ideal high-symmetry perovskite positions) presents a leading
antipolar instability (Figure~\ref{fig_gs}{\bf c}), but also
homogeneous polar instabilities with in-plane (Figure~\ref{fig_gs}{\bf
  d}) and out-of-plane (Figure~\ref{fig_gs}{\bf e}) polarizations.  It
is thus expected that an electric field along the stacking direction
could stabilize a monodomain configuration corresponding to
Figure~\ref{fig_gs}{\bf e}.

This type of dipole structure was predicted by
first-principles~\cite{aguado12}, second-principles~\cite{zubko16}
(like those used in this work), and phase-field~\cite{li17}
simulations, and was also experimentally
observed~\cite{yadav16,chen20}. Similar multidomain states have also
been reported for BaTiO$_3$/SrTiO$_{3}$ superlattices in the past
years~\cite{lisenkov07,estandia19,peng19}.

\begin{figure}
\includegraphics[scale=0.15]{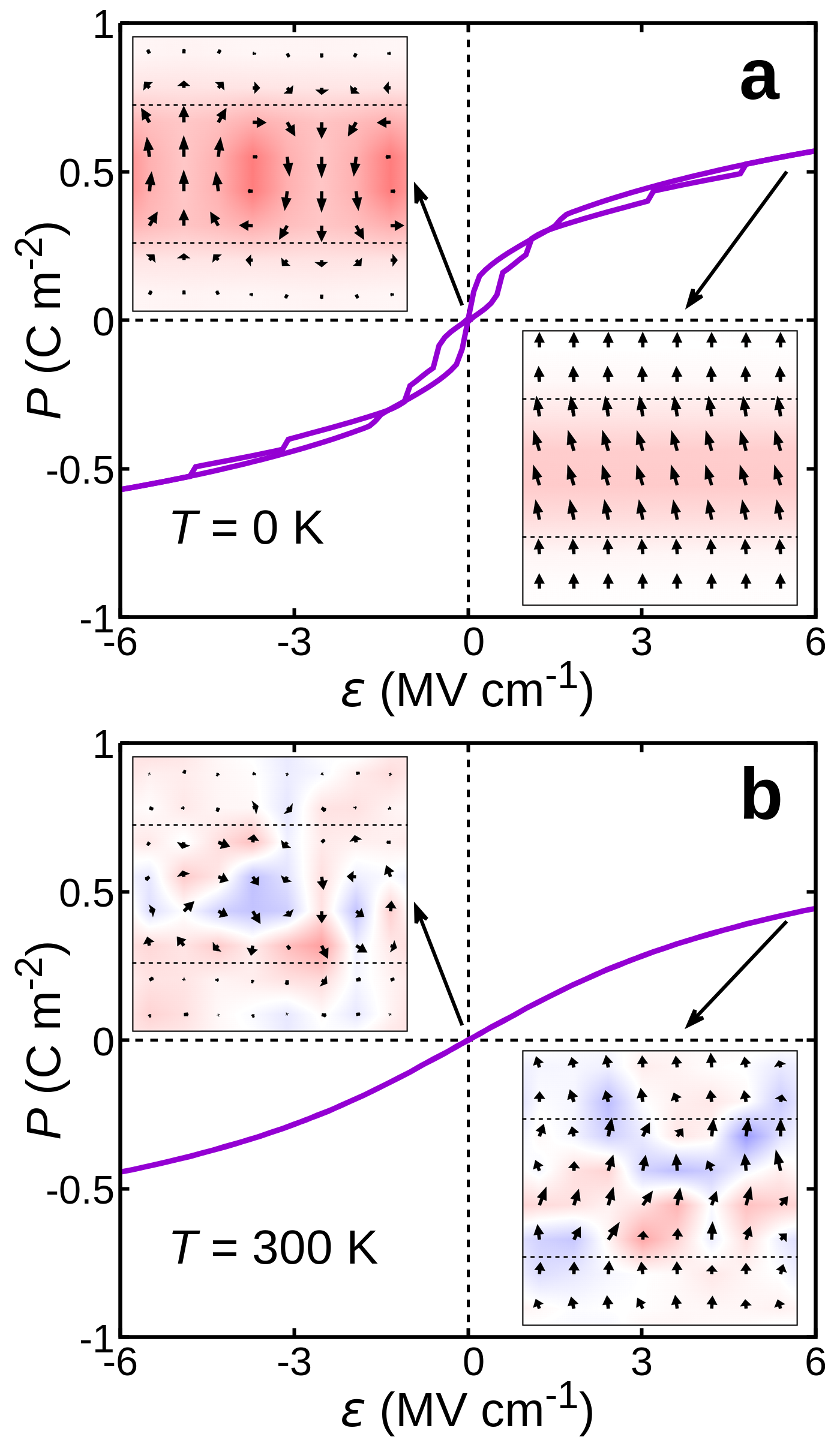}
	\caption{Representative $P$-$\varepsilon$ diagrams of PbTiO$_{3}$/SrTiO$_{3}$ superlattices at zero Kelvin ({\bf a}) and room temperature ({\bf b}).
	The insets in {\bf a} show the polarization state at zero field and at saturation.
	The insets in {\bf b} show representative snapshots of the system at zero field and at saturation. The colour scale for the polarization component perpendicular to the paper is that of Figure~\ref{fig_gs}{\bf a}.}
	\label{fig_PEtemp}
\end{figure}

We make use of Monte Carlo simulations under electric field (see
Methods) to compute polarization--electric field diagrams for these
systems. In Figure~\ref{fig_PEtemp}{\bf a} we show the response of the
(PbTiO$_3$)$_{4}$/(SrTiO$_{3}$)$_{4}$ superlattice at low temperatures
(strictly, 0~K). The material presents the mentioned antipolar state
at zero field, and undergoes a field-induced phase transition onto a
polar state for fields of a few MV~cm$^{-1}$. This transition occurs
in steps (corresponding to the switching of dipole columns in the
ferroelectric layer) and is slightly hysteretic. Our calculations thus
predict that these superlattices display antiferroelectric-like
behaviour.

In Figure~\ref{fig_PEtemp}{\bf b} we show the results of a similar
calculation at room temperature. It is apparent that the polarization
at high fields decreases with respect to the low temperature result,
and that the switching is smoother and non-hysteric. More importantly,
even if a static antipolar character at zero field is not apparent in
instantaneous configurations (see snapshot of the system in inset to
the left of Figure~\ref{fig_PEtemp}{\bf b}), the thermal average of
the polarization is clearly still zero. At large fields the
superlattice does present a clear polar order, as shown in the
representative system snapshot to the right of
Figure~\ref{fig_PEtemp}{\bf b}. This indicates that the
antiferroelectric-like behaviour is preserved at ambient conditions,
which makes PbTiO$_3$/SrTiO$_{3}$ superlattices a promising playground
for antiferroelectric capacitors.

We now investigate how the design parameters affect the performance of
the (PbTiO$_{3}$)$_{m}$/(SrTiO$_{3}$)$_{n}$ superlattices for energy
storage. To this end, we run high-throughput calculations of electric
field cycles up to 3.5~MV~cm$^{-1}$, with varying PbTiO$_{3}$ and
SrTiO$_{3}$ thicknesses (between 2 and 12, and between 2 and 20
perovskite unit cells, respectively), epitaxial strain $\eta$ (between
$-$3\% and $+$3\%, taking an SrTiO$_{3}$ substrate as the zero of
strain), and dielectric stiffness of the SrTiO$_{3}$ layer. This
stiffness can be controlled experimentally by chemical substitution,
and we model it by including an additional inter-atomic term (denoted
$h_{\rm STO}$) to favour or penalize polar distortions in SrTiO$_{3}$
(see Methods). In this way we obtain a database of more than 1250
$P$-$\varepsilon$ curves. We integrate the curves to obtain the stored
energy density as a function of the maximum applied field
($\varepsilon_{\mathrm{max}}$), and we compute the zero-field
susceptibility ($\chi_{0}$) and switching field
($\varepsilon_{\rm{sw}}$) of each $P$-$\varepsilon$ curve (see
Methods) to gain some insight into the results in terms of simple
physical descriptors.

\begin{figure}
	\includegraphics[scale=0.09]{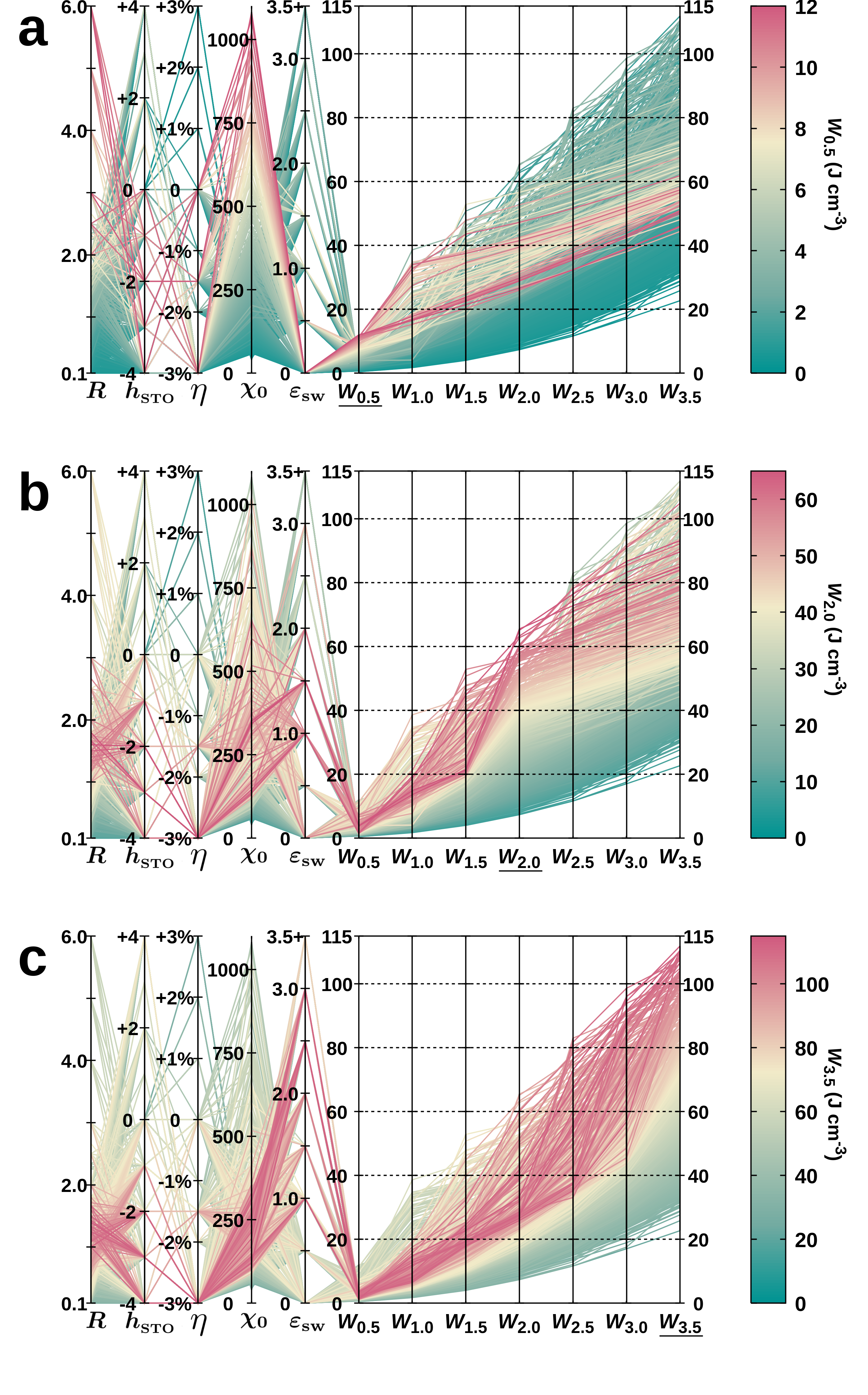}
	\caption{Parallel coordinates plots of the high-throughput data. The columns, from left to right, correspond respectively to PbTiO$_{3}$/SrTiO$_{3}$ ratio ($R$), modified SrTiO$_{3}$ stiffness ($h_{\rm{STO}}$), epitaxial strain ($\eta$), zero-field susceptibility ($\chi_{0}$), switching field ($\varepsilon_{\rm{sw}}$), and stored energy densities at different values of the applied electric field ($W_{0.5}$, $W_{1.0}$, $W_{1.5}$, $W_{2.0}$, $W_{2.5}$, $W_{3.0}$, and $W_{3.5}$). The lines are coloured according to $W_{0.5}$, $W_{2.0}$, and $W_{3.5}$, in panels {\bf a}, {\bf b}, and {\bf c}, respectively (corresponding colour scales to the right of each panel).}
	\label{fig_pcp}
\end{figure}

In order to identify correlations, we work with parallel coordinates
plots~\cite{inselberg85}. In these plots several vertical axes are
displayed in parallel, each representing one physical
descriptor. Every considered superlattice is represented by one line
in the plot. The lines are coloured according to the energy density at
a given maximum field (which is also represented in one of the
vertical axes), so that correlations can be visualized more easily. In
Figure~\ref{fig_pcp}{\bf a} we show the plot in which the colour scale
follows the stored energy density for a maximum applied field of
0.5~MV~cm$^{-1}$ ($W_{0.5}$). By visual inspection one can see that
the best superlattices for $\varepsilon_{\mathrm{max}}$ =
0.5~MV~cm$^{-1}$, shown in red, are those with the largest PbTiO$_{3}$
to SrTiO$_{3}$ thickness ratios $R=m/n$, the largest zero-field
susceptibilities, and the smallest switching fields. It is also clear
from the figure that a good performance at small applied fields
essentially implies a relatively poor performance at larger
fields. Even the less pronounced correlations of $W_{0.5}$ with
dielectrically softer SrTiO$_{3}$ (negative $h_{\rm{STO}}$) and
strains between 0 and $-$3\% are also clear from the figure. Similar
plots, in which the colour scale represents the stored energy density
at 2.0~MV~cm$^{-1}$ ($W_{2.0}$), and 3.5~MV~cm$^{-1}$ ($W_{3.5}$), are
shown in Figures~\ref{fig_pcp}{\bf b} and \ref{fig_pcp}{\bf c},
respectively. (For plots colour-coded according to the energy density
at intermediate fields see Supplementary Figure S1.)

We find a strong correlation between $\chi_{0}$ and the energy density
for small $\varepsilon_{\mathrm{max}}$ values (see
Fig.~\ref{fig_pcp}{\bf a}). Note that in this (linear) regime a large
$\chi_{0}$ implies a large polarization response for a small applied
field, which in turn translates into a large energy density. As the
field increases (and we move into the non-linear regime) we see that
the better performing superlattices show lower values of $\chi_{0}$. A
lower $\chi_{0}$ indicates a flatter initial slope in the
$P$-$\varepsilon$ diagram, which is beneficial for large enough
$\varepsilon_{\mathrm{max}}$ values (compare for instance the
performance of the red and green curves in
Figures~\ref{fig_explan}{\bf a} and~\ref{fig_explan}{\bf b}).

\begin{figure}
	\includegraphics[scale=0.19]{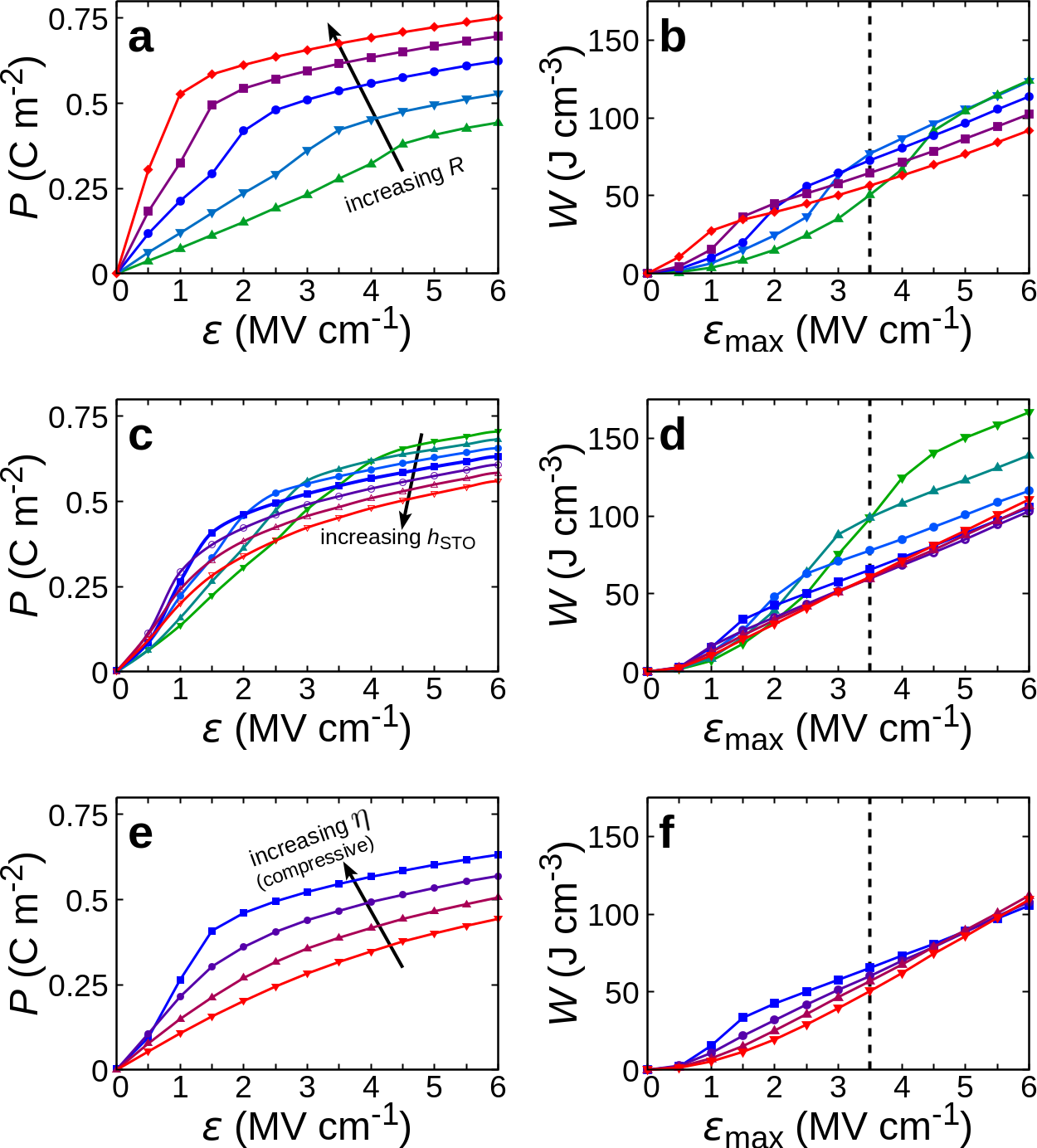}
	\caption{Effect of the different design parameters on the $P$-$\varepsilon$ loops (left column) and on the stored energy density (right column)	{\bf a} and {\bf b} show respectively how the $P$-$\varepsilon$ curves and stored energy density vary with the PbTiO$_{3}$ to SrTiO$_{3}$ ratio $R$. {\bf c} and {\bf d}: same as {\bf a} and {\bf b} for SrTiO$_{3}$ stiffness.	{\bf e} and {\bf f}: same as {\bf a} and {\bf b} for in-plane epitaxial strain $\eta$.}
	\label{fig_designparams}
\end{figure}

We observe that at low fields (Figure~\ref{fig_pcp}{\bf a}), a larger
$R$ ratio is correlated with a better performance, while at high
fields (Figure~\ref{fig_pcp}{\bf c}) the opposite holds true. To
better understand this behaviour, in
Figures~\ref{fig_designparams}{\bf a} and \ref{fig_designparams}{\bf
  b} we show how the variation of $R$ affects the polarization and
energy density of representative superlattices: a thicker
ferroelectric layer (or, equivalently, a thinner dielectric layer)
brings the system closer to the limit of a bulk ferroelectric
compound. This leads to a larger polarization in the switched state,
which in turn increases the energy density. However, it is also
apparent from these figures that this effect comes with a reduction of
the switching field, and that, in general, the optimal $R$ will depend
on the maximum applied field.

At low fields, the best performing superlattices tend to have
dielectrically-soft SrTiO$_{3}$ layers (negative $h_{\rm{STO}}$),
although many systems with unmodified stiffness (i.e. undoped
SrTiO$_{3}$, $h_{\rm{STO}}=0$) yield almost equally good
performances. A correlation between softened SrTiO$_{3}$ and high
performance is apparent at medium and large applied fields. We analyze
in detail the effect of varying $h_{\rm{STO}}$ in
Figures~\ref{fig_designparams}{\bf c} and~\ref{fig_designparams}{\bf
  d}. We see that a stiffer dielectric layer (positive $h_{\rm{STO}}$)
imposes a larger electrostatic penalty on the polar phase and hence
reduces the high-field polarization, which is detrimental for the
energy density. However, it also results in a decrease in the
switching field, so that (as it was the case for $R$) the optimal
dielectric stiffness depends on the maximum applied field (see
Figure~\ref{fig_designparams}{\bf d}).
 
A compressive epitaxial strain of up to $-$3\% is found to be
correlated with better performances in general. This is specially true
for intermediate and large maximum fields, while at low fields the
optimal strain window widens and includes unstrained
superlattices. The effect of varying the epitaxial strain is
illustrated in Figures~\ref{fig_designparams}{\bf e}
and~\ref{fig_designparams}{\bf f}. Compressive strain favours the
tetragonal distortion of PbTiO$_{3}$ and hence the alignment of its
polarization along the growth direction, yielding overall a larger
saturation polarization and increasing the energy density. At $-$3\%
strain we observe a flattening of the initial slope in the
$P$-$\varepsilon$ curve (i.e. a decrease in $\chi_{0}$) accompanied
with an increase in the switching field (from zero to a finite
value). This reflects the fact that the compressive strain yields a
more stable multidomain zero-field state with large local
polarizations, and therefore a higher energy barrier to escape out of
it. Ultimately this results in an increase in the energy density, as
shown in Figure~\ref{fig_designparams}{\bf f}. It thus seems that, at
least in the studied range of $\varepsilon_{\mathrm{max}}$,
compressive epitaxial strain always has a positive impact on the
stored energy density. This also points to the possibility of further
optimizing the superlattices by using a ferroelectric layer with a
larger bulk spontaneous polarization.

It is clear from Figure~\ref{fig_pcp} that the highest values of
$W_{0.5}$, $W_{2.0}$, and $W_{3.5}$ are respectively correlated with
switching fields of 0, just below 2.0~MV~cm$^{-1}$ and just below
3.5~MV~cm$^{-1}$. Indeed, we overall find that, given a maximum
applied field $\varepsilon_{\rm{max}}$, the stored energy density is
optimal for systems which have a switching field just below the
applied field ($\varepsilon_{\rm{sw}} \lessapprox
\varepsilon_{\rm{max}}$). A late switching is beneficial for the
stored energy since the area to the left of the $P$-$\varepsilon$
curve will be larger the later the polarization develops (e.g.
compare the red and green curves in Figures~\ref{fig_designparams}{\bf
  a} and~\ref{fig_designparams}{\bf b}). This conclusion is in line with the ideas put forward in Refs.~\onlinecite{burn72} and~\onlinecite{liu21}.

%\section*{Discussion}
\vspace{5mm}{\bf Discussion}

The PbTiO$_{3}$/SrTiO$_{3}$ superlattices studied in this work present
larger energy densities than most of the measured antiferroelectric
capacitors. At the highest field considered, $\varepsilon_{max}$ =
3.5~MV~cm$^{-1}$, our best superlattices store more than
110~J~cm$^{-3}$, which greatly exceeds the best results for
hafnia-based antiferroelectrics (less than
40~J~cm$^{-3}$)~\cite{park14}, or relaxor ferroelectric thin films
(almost 80~J~cm$^{-3}$)~\cite{kim20}, and is only surpassed by the
complex perovskite solid solution that holds the present record
(154~J~cm$^{-3}$)~\cite{peng15}, to our best knowledge.

If we focus on an intermediate field of $\varepsilon_{max}$ =
2.0~MV~cm$^{-1}$, we find maximum energy densities of 65~J~cm$^{-3}$,
which is larger than the largest value measured at that field in
complex perovskite solid solutions (almost
50~J~cm$^{-3}$)~\cite{li20a}, relaxor thin films (almost
30~J~cm$^{-3}$)~\cite{ma15}, or hafnia-based materials (around
10~J~cm$^{-3}$)~\cite{park14}.

For the lowest field here considered, $\varepsilon_{max}$ =
0.5~MV~cm$^{-1}$, the best PbTiO$_{3}$/SrTiO$_{3}$ superlattice stores
12~J~cm$^{-3}$, very close to the largest experimentally observed
value for perovskite solid solutions (12.6~J~cm$^{-3}$)~\cite{li20a},
and outperforming the reported relaxor thin films
(7.7~J~cm$^{-3}$)~\cite{ma09a} and hafnia-based materials (below
2~J~cm$^{-3}$)~\cite{park14}.

Let us also note that very high energy densities have recently been
predicted in AlN/ScN superlattices, up to 135~J~cm$^{-3}$ and
200~J~cm$^{-3}$ for very large fields of 5.0~MV~cm$^{-1}$ and
6.3~MV~cm$^{-1}$, respectively~\cite{jiang21}. Still, in the cited
work, these fields were rescaled by a factor of 1/3 to match an
experimental response curve, so a direct comparison to our results is
not possible (the fields actually considered in the simulations of
Ref.~\onlinecite{jiang21} were of the order of 15~MV~cm$^{-1}$, well beyond
typical breakdown fields). Along the same lines, in Ref.~\onlinecite{xu17}
lead-free perovskite solid solutions were predicted to display
energy-storage performances that exceed our present results; however,
electric fields were rescaled by a factor of 1/23 in that work, which
complicates a direct comparison.

While the second-principles models employed here have been shown to be
qualitatively correct, one may wonder about their quantitative
accuracy to reproduce experiments. We can get an idea by comparing our
results with the $P$-$\varepsilon$ curve of the
(PbTiO$_{3}$)$_{5}$/(SrTiO$_{3}$)$_{5}$ superlattice reported in Ref.~\onlinecite{zubko12}; there we find a polarization of around 0.2~C~m$^{-2}$
at 0.5~MV~cm$^{-1}$, while our simulations yield a polarization 3
times smaller for the same field. We attribute this to the fact that
our simulated SrTiO$_{3}$ layers are significantly stiffer than the
experimental ones~\cite{zubko16}. At any rate, this does not affect
the trends and basic quantitative results presented here. Indeed, we
could try to reproduce experimentally our simulated superlattices by
considering Zr-doped SrTiO$_{3}$, which should result in relatively
stiff dielectric layers.

In conclusion, we have computed the room-temperature energy storage
capabilities of over 1000 PbTiO$_{3}$/SrTiO$_{3}$ superlattices with
different defining parameters. This high-throughput approach (possible
thanks to second-principles methods) allows us to identify optimal
conditions, predicting that these systems outperform most of the
reported antiferroelectric capacitors in a wide range of applied
fields. The best materials consistently present a switching field just
below the maximum applied field, indicating that tuning this variable
is key to improving energy-storing performance. Moreover, we find that
these superlattices can be tailored to address specific needs by means
of strain, layer thicknesses and dielectric stiffness, depending on the
available or desired maximum applied fields. Hence, our results
indicate that electrostatically engineered ferroelectric/paraelectric
superlattices are promising materials for applications in pulsed power
technologies.

%\noindent {\bf Supplementary Material} accompanies this paper at {\small {\tt http://www.scienceadvances.org/}}.

%\section*{Methods}
\vspace{5mm}{\bf Methods}

We run second-principles simulations as implemented in the \textsc{SCALE-UP} code~\cite{wojdel13,garcia16,escorihuela17}. The models for the superlattices are derived from models for bulk SrTiO$_{3}$ and bulk PbTiO$_{3}$ which have been used in previous works~\cite{wojdel13,wojdel14,seijas17} and give correct descriptions of the lattice dynamical properties of both compounds. Then, as described also in Ref.~\onlinecite{das19}, the interactions involving interfacial atoms in the superlattices are taken as the numerical average of the corresponding interactions in PbTiO$_{3}$ and SrTiO$_{3}$. 
In order to reproduce the correct long-range electrostatic behaviour, an effective dielectric tensor $\epsilon^{\infty, \rm{SL}}$ is used for the superlattice. Along the growth direction of the superlattice the system is considered as capacitors in series, so that the inverse of the diagonal component of the electrostatic tensor along the growth direction, $\epsilon^{\infty, \rm{SL}}_{zz}$ is taken as the weighted sum of the inverses of the corresponding tensor elements for bulk PbTiO$_{3}$ ($\epsilon^{\infty, \rm{PTO}}_{zz}$) and SrTiO$_{3}$ ($\epsilon^{\infty, \rm{PTO}}_{zz}$) as obtained from first principles, where the weights ($t_{\rm{PTO}}$ and $t_{\rm{STO}}$) are the relative thicknesses of the respective layers of PbTiO$_{3}$ and SrTiO$_{3}$:
\begin{equation}
(\epsilon^{\infty, \rm{SL}}_{zz})^{-1}=t_{\rm{PTO}}(\epsilon^{\infty,\rm{PTO}}_{zz})^{-1}+t_{\rm{STO}}(\epsilon^{\infty,\rm{STO}}_{zz})^{-1}
\end{equation}
Analogously, for the in-plane components of the electrostatic tensor the layers are considered as capacitors in parallel, resulting in an effective in-plane electrostatic tensor given by:
\begin{equation}
\epsilon^{\infty, \rm{SL}}_{ii}=t_{\rm{PTO}}\epsilon^{\infty,\rm{PTO}}_{ii}+t_{\rm{STO}}\epsilon^{\infty,\rm{STO}}_{ii} \text{for $i=x,y$}
\end{equation}
Finally, and in order to recover the correct bulk limits, the Born effective charges ($Z^{*,\rm{SL}}_{\alpha}$, where $\alpha$ runs through the atoms) within each layer $j$ are rescaled by 
$\sqrt{\epsilon^{\infty,\rm{SL}}/\epsilon^{\infty,j}}$ where $j$= PTO, STO. For the Ti and O atoms at the interfaces the Born effective charges are renormalized as follows:
$Z^{*}_{\alpha}=\frac{1}{2}\sqrt{\epsilon^{\infty,\rm{SL}}/\epsilon^{\infty,j}}Z^{*,\rm{PTO}}_{\alpha}+\frac{1}{2}\sqrt{\epsilon^{\infty,\rm{SL}}/\epsilon^{\infty,j}}Z^{*,\rm{STO}}_{\alpha}$
where $Z^{*,\rm{PTO}}_{\alpha}$ and $Z^{*,\rm{STO}}_{\alpha}$ are the Born effective charges of atom $\alpha$ in bulk PbTiO$_{3}$ and SrTiO$_{3}$, respectively, and $\alpha$= Ti, O at the interface.

The second-principles parameters of both materials are fitted from density functional theory calculations at a hydrostatic pressure of $-$11.2~GPa to correct for the underestimation due to the local density approximation of the cubic lattice constant that is taken as the reference structure. 

The dielectric stiffness of the SrTiO$_{3}$ layer is modified by adding an extra interatomic term to the superlattice model, with the representative term (Ti$_{z}$-O$_{z}$)$^{2}$, that only affects Ti and O atoms in the SrTiO$_{3}$ subsystem. (Note that this expression is merely the representation of a symmetry-adapted term~\cite{wojdel13} but also affects polar distortions in SrTiO$_{3}$ along the $x$ and $y$ directions.) In this way, a positive coefficient translates into an additional energy cost of polarizing the SrTiO$_{3}$ layer (and hence the superlattice) in the growth direction. In the high-throughput calculations the coefficient for this term, $h_{\rm{STO}}$, is varied between $-$4 meV \AA$^{-2}$ and $+$4 meV \AA$^{-2}$.

For the ground state calculations and the simulation of the zero Kelvin $P$-$\varepsilon$ diagrams we run Monte Carlo simulated annealings for 30000 steps, with an initial temperature of 10~K and an annealing rate of 0.9975. To simulate electric field cycles, the studied electric field range (0 to 6~MV~cm$^{-1}$ in Figures~\ref{fig_PEtemp} and \ref{fig_designparams}, 0 to 3.5~MV~cm$^{-1}$ in the high-throughput calculations) is divided in increments of equally length (of 0.2 MV~cm$^{-1}$ everywhere, except for the high-throughput calculations for which an electric field step of 0.5~MV~cm$^{-1}$ was employed).
For each value of the electric field a Monte Carlo simulation is performed sequentially (an annealing for the zero Kelvin diagrams, and a constant temperature Monte Carlo simulation for the finite temperature simulations), using as initial configuration that of the previous step in the electric field ramp. To generate the finite temperature $P$-$\varepsilon$ curves we run Monte Carlo simulations at constant temperature for 30000 steps at each value of the electric field, which we find to be enough to show converged results. The averages of the polarization at each value of the electric field are taken disregarding the initial 5000 steps of each simulation, to allow for thermalization.

The high-throughput calculations are performed in a simulation cell of 8$\times$2$\times$1, where the unitary cell is defined as a 1$\times$1 perovskite unit cells in the $xy$ plane and a full superlattice period in the third direction. We check the convergence of our calculations with respect to the simulation cell. To this end, we compare the $P$-$\varepsilon$ curves of the (PbTiO$_3$)$_{4}$/(SrTiO$_{3}$)$_{4}$ in simulation cells of 8$\times$2$\times$1, 8$\times$8$\times$1 and 12$\times$12$\times$1, both under strains of 0\% and $-$3\% (see Supplementary Figure S2). We find that the results for the 8$\times$8$\times$1 cell are very well converged, since they are essentially identical to those of the 12$\times$12$\times$1 cell. The 8$\times$2$\times$1 cell also yields very well converged results under no strain. Under compressive strain the switching field (inflection point of the curve) becomes finite and the results for the 8$\times$2$\times$1 cell are not so well converged around $\varepsilon_{\mathrm{sw}}$. Still, the effect is not very large, and the polarization is underestimated at lower fields, then overestimated at intermediate fields, and finally well converged for high fields. Overall, the effect in the stored energy density is not large (specially for fields above $\varepsilon_{\mathrm{sw}}$).

For a given electric field ramp the zero-field susceptibility $\chi_0$ is computed using finite differences. The switching field is defined as the inflection point in the $P$-$\varepsilon$ curves. In order to estimate the $\varepsilon_{\rm{sw}}$ in the $P$-$\varepsilon$ curves we compute their second derivatives using central finite differences and we set $\varepsilon_{\rm{sw}}$ to the largest field for which the second derivative is positive (positive curvature). In the cases where the curvature of the $P$-$\varepsilon$ curve is found to be negative for all the studied electric fields, $\varepsilon_{\rm{sw}}$ is set to zero. When the curvature is found to be positive for the full range of electric field studied (up to 3.5~MV~cm$^{-1}$), since the large field behaviour has to be that of a saturating polarization with negative polarization, we set the $\varepsilon_{\rm{sw}}$ to be 3.5~MV~cm$^{-1}$ or more (3.5+ in Figures~\ref{fig_pcp} and S1).

The stored energy density at each value of the field is computed by trapezoid integration of the $P$-$\varepsilon$ over the $P$ axis. 

In the high-throughput calculations, we run a full charge-discharge cycle (from zero to 3.5~MV~cm$^{-1}$, then back to zero field) for several sets of design parameters (more than 50) to test for possible hysteresis, finding that none of the systems presented hysteric behaviour.

%{\bf Acknowledgements:} 
\noindent {\bf Acknowledgements:} Work funded by the Luxembourg National Research Fund through project C18/MS/12705883 ``REFOX''.\\
%\noindent {\bf Funding:} Work funded by the Luxembourg National Research Fund through project C18/MS/12705883 ``REFOX''.\\
%{\bf Author Contributions.}
\noindent {\bf Author Contributions.} J.\'I. conceived the research. H.A. performed the calculations and created the figures. All authors contributed to the discussion and analysis of the results. The manuscript was written by H.A. and J.\'I., with contributions from N.S.F.\\
\noindent{\bf Competing Interests.} The authors declare that they have no competing financial interests.\\
\noindent{\bf Data availability:} Data related to this work is available from the authors upon reasonable request.

\end{document}